\documentclass[twocolumn,showpacs,preprintnumbers,amsmath,amssymb,prl,superscriptaddress]{revtex4}

\usepackage{latexsym}
\usepackage{epsfig}
\usepackage{bm}
\usepackage{wasysym}

\usepackage{graphicx}% Include figure files
\usepackage{dcolumn}% Align table columns on decimal point
\usepackage{hyperref}

\hypersetup{
 pdfauthor = {Markus B\"uscher, FZ J\"ulich},
 pdftitle = {Non-Rayleigh Decays of Thin Cryogenic Jets},
 pdfsubject = {Manuscript for PRL},
 pdfkeywords = {},
 pdfcreator = {LaTeX with hyperref package},
 pdfproducer = {dvipdf}}

\begin{document}

\title{\boldmath Dynamics of Cryogenic Jets:\\Non-Rayleigh Breakup and
  Onset of Non-Axisymmetric Motions}

\author{A.V.~Boukharov}
\affiliation{Moscow Power Engineering Institute, Ul. Krasno-Kazarmennaja 13, 
                     117218 Moscow, Russia}%
\author{M.\,B\"uscher}\email[E-mail: ]{m.buescher@fz-juelich.de}
\affiliation{Institut f\"ur Kernphysik, Forschungszentrum J\"ulich, 
                     52425 J\"ulich, Germany}%
\author{A.S.~Gerasimov}
\affiliation{\mbox{Institute for Theoretical and Experimental Physics, B. Cheremushkinskaya 25, 
                     117218 Moscow, Russia}}
\author{V.D.~Chernetsky}
\affiliation{\mbox{Institute for Theoretical and Experimental Physics, B. Cheremushkinskaya 25, 
                     117218 Moscow, Russia}}
\author{P.V.~Fedorets}
\affiliation{Institut f\"ur Kernphysik, Forschungszentrum J\"ulich, 
                     52425 J\"ulich, Germany}%
\affiliation{\mbox{Institute for Theoretical and Experimental Physics, B. Cheremushkinskaya 25, 
                     117218 Moscow, Russia}}
\author{I.N.~Maryshev}
\affiliation{Moscow Power Engineering Institute, Ul. Krasno-Kazarmennaja 13, 
                     117218 Moscow, Russia}%
\author{A.A.~Semenov}
\affiliation{Moscow Power Engineering Institute, Ul. Krasno-Kazarmennaja 13, 
                     117218 Moscow, Russia}%
\author{A.F.~Ginevskii}
\affiliation{Moscow Power Engineering Institute, Ul. Krasno-Kazarmennaja 13, 
                     117218 Moscow, Russia}%

\date{\today}

\begin{abstract}
  We report development of generators for periodic, satellite-free
  fluxes of mono-disperse drops with diameters down to $10\,\mu$m from
  cryogenic liquids like H$_2$, N$_2$, Ar and Xe (and, as reference
  fluid, water). While the breakup of water jets can well be described
  by Rayleigh's linear theory, we find jet regimes for H$_2$ and N$_2$
  which reveal deviations from this behavior.  Thus, Rayleigh's theory
  is inappropriate for thin jets that exchange energy and/or mass with
  the surrounding medium.  Moreover, at high evaporation rates, axial
  symmetry of the dynamics is lost. When the drops pass into vacuum,
  frozen pellets form due to surface evaporation. The narrow width of
  the pellet flux paves the way towards various industrial and
  scientific applications.
\end{abstract}
\maketitle

Drop formation from incompressible liquids is ubiquitous in daily life
and technology, and such processes have been at the focus of
scientific investigations for almost two
centuries~\cite{1,2,3,chandrasekhar,4,5}.  Even the simple case of
jets emanating from a vibrating nozzle leads to a broad spectrum of
drop sizes, and the existence of satellite drops~\cite{6}.  However,
for technologies like ink-jet printing~\cite{7}, for laser-plasma UV
sources~\cite{10}, for compact laser-based particle
accelerators~\cite{11}, accelerator experiments~\cite{12}, or for
space operations~\cite{16}, the drop sizes and rates must be highly
homogeneous.

The disintegration of a slow cylindrical jet is caused by the growth
of perturbations initiated at the ejection nozzle. An axially
symmetric perturbation leads to jet breakup when its amplitude becomes
equal to the jet radius. This domain is commonly denoted as the
Rayleigh regime. In 1878 Lord Rayleigh described fluids in the limit
of zero viscosity~\cite{2}, while his 1892 paper~\cite{3} concerns the
opposite limit where inertial effects are negligible compared to
viscous ones. Rayleigh's work was extended by
Chandrasekhar~\cite{chandrasekhar} to arbitrary viscosities.  It has
been shown~\cite{6} that the measurement of breakup lengths, {\em
  e.g.\/} for the forced breakup of water jets with precise control of
the nozzle-perturbation frequency, allows for precise tests of linear
theories, although the final stage of capillary pinching is
non-linear~\cite{non-linear}.  

Experimentally, the selection of a well defined initial
surface-perturbation amplitude and a certain frequency is commonly
achieved by using piezoelectric transducers.  Most of the existing
data have been obtained for water (or water-based ink) and fluids of
higher viscosities, with jets emanating into air at normal pressure.

In this letter we report on first quantitative measurements of the
breakup of cryogenic and water jets, injected into the same gaseous
substance close to triple-point (TP) conditions (63~K and 125~mbar for
N$_2$, 14~K and 70~mbar for H$_2$). In doing so we can widen the
experimentally accessible parameter space, while keeping the Reynolds
number $\mathsf{Re} {=} ( \rho\cdot v_{\mathrm{jet}} \cdot R_0)/\mu$
(which represents the relative importance of inertial and viscous
forces) and the Ohnesorge number $\mathsf{Oh} {=} \mu/(\sigma\cdot
\rho \cdot R_0)^{1/2}$ (viscous forces and surface tension) in the
same range as for the existing water data. $\rho$ ($\mu$, $\sigma$)
denote the fluid density (viscosity, surface tension), and
$v_{\mathrm{jet}}$ ($R_0$) the jet velocity (initial jet radius).  At
TP pressures and jet velocities of a few m/s, aerodynamic interactions
with the ambient medium are small and, thus, we are close to the ideal
case of an inviscid jet breaking up in vacuum and our data allow for
clean tests of Rayleigh's theory.

\begin{figure}[htb]
\scalebox{0.55}[0.5]
  {\includegraphics{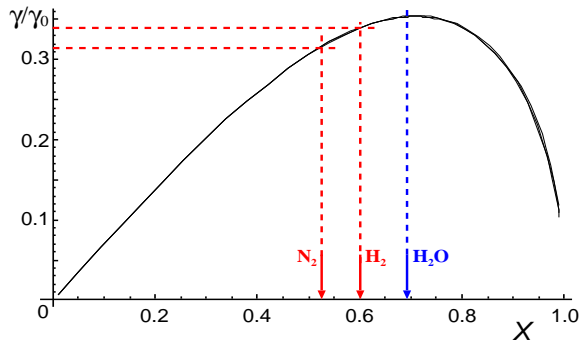}}
  \vspace*{-4mm}
  \caption{\label{fig:gamma} Dimensionless growth rate
    $\gamma/\gamma_0$ as a function of the reduced wave number $X$ in
    linear theory. For N$_2$ and H$_2$ two (practically identical)
    curves have been calculated with Eq.(\ref{eq:2}) using the
    $\mathsf{Oh}$ values from Table~\ref{tab:1}. For water a typical
    Ohnesorge number $\mathsf{Oh}{=} 0.02$ has been chosen, since
    $\gamma/\gamma_0$ behaves similarly at $\mathsf{Oh}< 0.1$.  The
    arrows indicate the $X$-values for which we have observed
    satellite-free drop production.}
\end{figure}

According to the linear theory, a small surface perturbation $\delta$
(measured in units of $R_0$) of a jet moving into $z$-direction,
imposed at the nozzle exit ($z{=}0$) with frequency $f$, grows
exponentially towards the jet axis~\cite{2,20}. Then, the minimum
(over time) necking of the jet $R(z)$ at a certain distance $z$ from
the nozzle is
\begin{equation}
  \frac{R(z)}{R_0} = 
     1 - \delta \exp{\left(\frac{\gamma/\gamma_0}{\mathsf{Oh}\cdot\mathsf{Re}}
                      \cdot \frac{z}{R_0}\right)}\ .
  \label{eq:1}
\end{equation}
The dimensionless growth rate $\gamma/\gamma_0$ depends on the
wavelength $\lambda$ of the perturbation and can be expressed in terms
of a reduced wave number $X {=} 2 \pi\cdot R_0/\lambda = 2\pi\cdot
R_0\cdot f/v_{\mathrm{jet}}$ for $X<1$ and
$\mathsf{Re}\cdot\mathsf{Oh} \gg 1$:
\begin{equation}
  \frac{\gamma}{\gamma_0} = -\frac{3}{2}\mathsf{Oh}X^2 + 
        \sqrt{\frac{9}{4}\mathsf{Oh}^2X^4 + \frac{1}{2}X^2(1-X^2)}
  \label{eq:2}
\end{equation}
with $\gamma_0 {=} \mathsf{Oh}\cdot (\sigma / \mu R_0)$.
Figure~\ref{fig:gamma} shows a plot of Eq.(\ref{eq:2}) for water,
N$_2$ and H$_2$ jets; linear theory predicts a practically identical
$X$ dependence of $\gamma/\gamma_0$ for all three liquids. If
$\mathsf{Re}\cdot\mathsf{Oh}$ approaches unity, the theory tends to
underestimate $\gamma/\gamma_0$, see {\em e.g.\/} the discussion by
Kalaaji et al.~\cite{6}.

Since a jet breaks into drops for $R(z){=}0$, the jet length
$L_{\mathrm{jet}}$ results from Eq.(\ref{eq:1}) as:
\begin{equation}
  L_{\mathrm{jet}} = 
    \frac{R_0}{\gamma/\gamma_0}\ \mathsf{Re}{\cdot}\mathsf{Oh}\ \ln{(1/\delta)}=
    \frac{v_{\mathrm{jet}}}{\gamma/\gamma_0}\ 
                   \sqrt{\frac{\rho R_0^3}{\sigma}} 
                         \ \ln{(1/\delta)}
  \label{eq:3}
\end{equation}
Relations (\ref{eq:1})--(\ref{eq:3}) have been experimentally verified
in numerous studies with water or viscous fluids~\cite{6,17}.
Satellite-free and mono-disperse water-drop production has also been
reported: for not too large values of $\delta$ ($\lesssim 0.01$) it
only occurs at maximum values of $\gamma/\gamma_0$, {\em i.e.\/} at
$X_{\mathrm{max}}{=}0.69$~\cite{18,24}.  

\begin{figure}[htb]
\scalebox{0.85}[0.80]
  {\includegraphics[width=\linewidth]{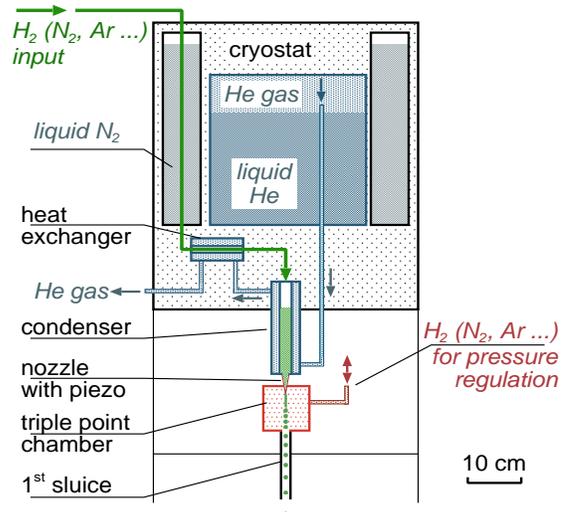}}
\vspace*{-2mm}
\caption{\label{fig:target} (Color online) Sketch of the drop
  generator. The cryostat (total height $\sim$0.8~m, not drawn to
  scale) comprises the cooling liquids (N$_2$ and He) as well as the
  heat exchanger and the upper part of the condenser. These are cooled
  by a stream of evaporated He gas. The condenser houses a few cm$^3$
  of the cryogenic liquid which is driven through a nozzle into the
  TPC by overpressure. Constant pressure in the TPC is maintained by
  an auxiliary gas feeding.}
\end{figure}

Quantitative studies of forced jet break-up depend crucially on the
suppression of unwanted nozzle vibrations which can, {\em e.g.}, be
caused by cold head units~\cite{27}. Thus our drop
generator~\cite{patent} is surrounded by a cryostat with baths of
liquid N$_2$ and He, see Fig.~\ref{fig:target}.  Liquefaction of H$_2$
is achieved in three stages: first cooling with liquid N$_2$, further
cooling in a heat exchanger by evaporated He, and final cooling in the
condenser by cold He gas. For production of, {\em e.g.}, N$_2$ and Ar
jets, cooling with liquid N$_2$ is sufficient while for water we use a
dedicated generator without cooling cryostat.  The temperatures along
the main gas and liquid flows are controlled with an accuracy of
0.1~K.  From the condenser, the liquid passes through the vibrating
nozzle and reaches the triple-point chamber (TPC) where the jet
disintegrates into drops.  The liquid temperature in the nozzle is
maintained slightly below the boiling point and, according to
simulation calculations, the temperature decrease in the jet until
break-up from heat exchange in the TPC is below 0.2~K (cf.\
Table~\ref{tab:1}).  Therefore, the generation of semi-solid domains
can be excluded.  The temperatures and pressures in the TPC are kept
over several hours within $\pm 0.2$~K ($\pm 0.2$~mbar) at the nominal
values.

\begin{table}[b]
  \caption{\label{tab:1} Parameters of the jets from Fig.~\ref{fig:H2-N2}. 
    $\gamma/\gamma_0$ is calculated with Eq.(\ref{eq:2}), 
    the jet length $L_{\mathrm{jet}}^{\mathrm{Rayleigh}}$ 
    with Eq.(\ref{eq:3}).}
  \begin{ruledtabular}
   \begin{tabular}{lcc}
             & N$_2$ & H$_2$ \\
     \hline
     Density $\rho$ (kg/m$^3$) \cite{25,26}& 823.7 & 64.04\\
     Static surface tension $\sigma$ (N/m) \cite{25,26}&0.0094&0.0025\\
     Viscosity $\mu$ (mPa\, s) \cite{25,26} & 0.2113 & 0.0183\\
     TPC pressure (mbar) & 300 & 130\\
     TPC temperature (K) & 74  & 17\\
     Jet temperature (K) & 77.2--77.0  & 20.0--19.8\\
     Nozzle frequency $f$ (kHz) & 26 & 38 \\
     Jet diameter $2R_0$ ($\mu$m) & 17 & 12 \\
     Jet velocity $v_{\mathrm{jet}}$ (m/s) & 2.6 & 2.4\\
     Reduced wave number $X$ &  0.53 & 0.60 \\
     Reynolds number $\mathsf{Re}$ & 84 & 50 \\
     Ohnesorge number $\mathsf{Oh}$ & 0.026 & 0.019 \\
     $\mathsf{Re}\cdot\mathsf{Oh}$ & 2.2 & 1.0 \\
     Growth rate $\gamma/\gamma_0$ & 0.31 & 0.34 \\
     Initial perturbation $\delta$ & 0.02 & 0.02\\
     Expected jet length $L_{\mathrm{jet}}^{\mathrm{Rayleigh}}$ ($\mu$m) 
                        & 240 & 70 \\
     Measured jet length $L_{\mathrm{jet}}$ ($\mu$m)& 310 & 290 \\ 
   \end{tabular}
 \end{ruledtabular}
\end{table}

Two nozzle types have been utilized, glass nozzles in brass housings
and such made from stainless steel. The former, with inner diameters
of 12--$30\, \mu$m at the nozzle tip, have the advantage of a smooth
internal surface and allow one to look inside the channel during
operation. The 16--$30\, \mu$m steel nozzles offer high shape
reproducibility and smaller length-to-diameter ratios of the holes,
allowing operation with lower jet-driving pressures. The results
presented here have been obtained with glass nozzles and pressures of
$\sim$0.4--0.9 bar for H$_2$ and $\sim$1.0--1.5 bar for N$_2$.  Higher
pressures lead to higher jet velocities. The piezo-electric
transducer allows us to excite sinusoidal nozzle vibrations in the
range $f{=}1$--60\,kHz.

The jet decay has been observed with two perpendicularly placed CCD
cameras and strobe lamps that are triggered synchronously to the
nozzle, with a 10 times lower frequency, and a flash duration of
$\sim$$1.5\,\mu$s.  The jet diameter $2R_0$ has been assumed to be
equal to the nozzle diameter which is correct to better than $\pm
10$\%~\cite{6}. This is confirmed by our measurement of jet diameters
with the CCD cameras.  $v_{\mathrm{jet}}$ has been determined by
measuring the wave-propagation velocity along the jet surface. The
initial amplitude $\delta$ has been derived from reference
measurements of water jet lengths using Eq.(\ref{eq:3}), as well as
from direct amplitude measurements with an Michelson interferometer.
See Ref.~\cite{17} for further details of the jet diagnostics.

\begin{figure}[b]
\scalebox{0.79}[0.78]
  {\includegraphics[width=\linewidth]{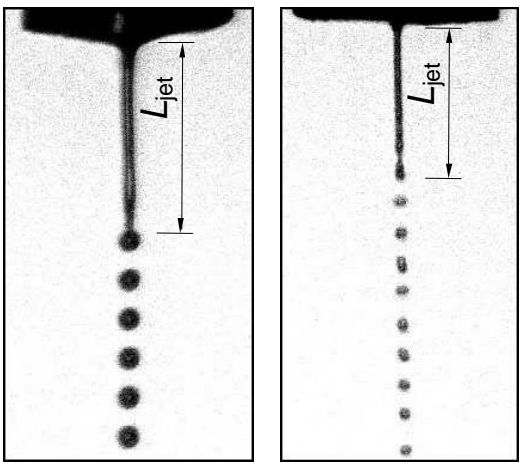}}
  \vspace*{-3mm}
  \caption{\label{fig:H2-N2} First observation of satellite-free and
    monodisperse disintegration of N$_2$ (left) and H$_2$ (right)
    jets. In the upper edge of the photos the tip of the vibrating
    nozzle can be seen.}
\end{figure}

Figure~\ref{fig:H2-N2} shows the breakup of N$_2$ and H$_2$ jets; for
certain choices of the jet and TPC parameters (see Table~\ref{tab:1}),
satellite-free and mono-disperse drop production is achieved.
Astonishingly, we find it at $X {=} 0.53$ and $X {=} 0.60$,
respectively, {\em i.e.\/} for different wave numbers as compared to
water. Since this is off the maximum from Fig.~\ref{fig:gamma}, it
suggests an $X$ dependence of $\gamma/\gamma_0$ different from
Eq.(\ref{eq:2}). In order to further test a possible deviation from
Rayleigh-like behavior, we have checked the validity of
Eqs.~(\ref{eq:1})--(\ref{eq:3}) in more detail.

The last two lines of Table~\ref{tab:1} compare the measured and
predicted jet-decay lengths; it is seen that the cryogenic jets are
much longer than expected. This cannot be explained by a variation of
the parameters that enter Eq.(\ref{eq:3}). For example, an unphysical
value of $\delta {=} 10^{-8}$ would be required to yield the large
decay length of the H$_2$ jet.  The dynamic surface tension can differ
from the static values for $\sigma$ quoted in Table~\ref{tab:1}, it
had to be, however, 20 (2) times smaller to explain the observed H$_2$
(N$_2$) jet lengths which seems unreasonable.

\begin{figure}[b]
\scalebox{1.0}[1.0]
  {\includegraphics[width=\linewidth]{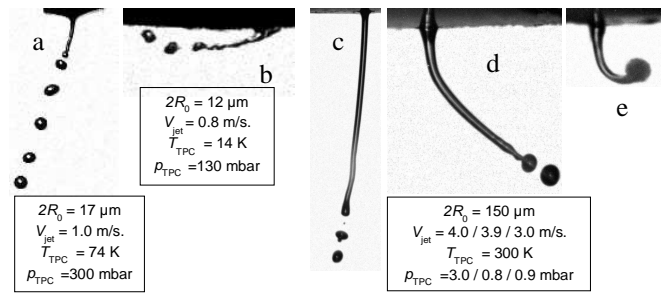}}
  \vspace*{-6mm}
  \caption{\label{fig:bending}First observation of non-axisymmetric
    N$_2$ (a), H$_2$ (b) and water (c--e) jets.}
\end{figure}

An explanation of the obvious violation of Rayleigh's theory might be
disturbances like temperature, pressure and surface-tension variations
as well as liquid-gas transitions.  It has, {\em e.g.}, been
predicted~\cite{evaporation} that evaporation and condensation
decrease $L_{\mathrm{jet}}$ (which seems to contradict the large
measured $L_{\mathrm{jet}}$ values from Table~\ref{tab:1}) and, for
large evaporation rates, satellite production can be suppressed.  One
may presume that for our jets evaporation is strong since in the TPC
jets and ambient gas coexist close to the vapor-pressure curve.  In
order to test the effect of evaporation we have enhanced it by
reducing the jet velocities or the TPC pressure (keeping the other
parameters constant). This leads to a phenomenon, to our knowledge not
reported until now. The jets slowly move from their vertical flux
direction to one off the nozzle symmetry axis while preserving their
smooth surfaces. The direction they choose seems to be random.
However, once a jet has arrived at a certain stable configuration it
may stay there for seconds; only at very small jet-bending radii are
frequent abrupt changes of the jet directions observed. As an example
we show in Fig.~\ref{fig:bending} N$_2$, H$_2$ and water jets. Even
back-bending is observed in some cases.  Systematic studies reveal
larger jet deflections from the vertical axis with decreasing jet
diameter and velocity or TPC pressure. We note that asymmetric heating
has been employed by another group to deflect liquid
microjets~\cite{deflection}, however, we report spontaneous deflection
of symmetrically produced jets.

When the drops leave the TPC through a $1^{\mathrm{st}}$ sluice into a
subsequent chamber ($p {=} o(10^{-2})$~mbar), they freeze to pellets
due to strong surface evaporation, and are accelerated by the gas flow
from the TPC.  Then the pellets pass a $2^{\mathrm{nd}}$ sluice and a
$2^{\mathrm{nd}}$ chamber ($o(10^{-4})$~mbar), and finally reach a
dummy scattering chamber through a thin tube ($\diameter \sim 2$~cm),
which is located 1.2 m downstream of the TPC.  In order to minimize
turbulences of the gas flow in the $1^{\mathrm{st}}$ sluice (and,
thus, distortions of the produced pellets) it has a circular cross
section with a radius that decreases exponentially to $600\, \mu$m in
flight direction.  For pellet observation two CCD cameras have been
positioned at the outlet of the $1^\mathrm{st}$ sluice or at the dummy
chamber. Stable H$_2$ and N$_2$ pellet production with diameters
20--$40\,\mu$m (the pellet diameter is about two nozzle diameters) has
been observed.  Deviations from the mean pellet diameter are below 1\%
(10\%) over periods of few seconds (hours).  The average velocity of
$30\,\mu$m pellets amounts to $\sim$70~m/s.  The radial displacement
of the pellets from their nominal flight path in the dummy chamber of
about $\pm 200\,\mu$m has been extrapolated from the angular pellet
distributions measured behind the $1^{\mathrm{st}}$ sluice.  This
value is dominated by our experimental resolution.  The small
pellet-beam divergence is a consequence of the extremely regular jet
breakup in the TPC (Fig.~\ref{fig:H2-N2}).

If used as an internal target at a storage ring, our device minimizes
unwanted gas loads to the accelerator and allows for an effective
target thickness of a few $10^{15}$~atoms/cm$^2$. Since the tube
connecting the generator with the interaction zone is rather narrow,
detectors can be placed close to the interaction point in a nearly
$4\pi$ configuration. Such a geometry is also advantageous for placing
collector mirrors in laser-plasma UV sources~\cite{10}.  In addition,
the large separation of the hot and radiating plasma from the jet
generator makes such a set-up especially favorable.

In this letter we present first data on the mono-disperse breakup of
cryogenic jets. We find that the jets are significantly more resistant
to breakup than predicted by Rayleigh's theory. We suggest that a
reason for this discrepancy is the influence of evaporation effects.
This interpretation is supported by the observation of non-axially
symmetric jets.  This new jet mode --- for which so far not even a
rudimentary theoretical explanation has been formulated --- sets in
when one decreases the ambient pressure very close to the triple-point
values.  For a complete theoretical delineation of drop formation
processes from thin cryogenic jets, further systematic data, like on
the $X$ dependence of $\gamma/\gamma_0$ and the pressure dependence of
$L_{\mathrm{jet}}$, are required.

\begin{acknowledgments}
  This work has been supported by BMBF (WTZ), COSY-FFE, DFG, EU (FP6),
  INTAS, ISTC, RFFI (07-08-00747a, 08-08-91950-NNIO\_a), RMS, and the
  PANDA collaboration. The Russian team members thank for the
  hospitality at the FZJ, in particular for the support by
  H.Str\"oher. We acknowledge discussions with H.Calen, C.Ekstr\"om
  and \"O.Nordhage.
\end{acknowledgments}


\begin{thebibliography}{99}

\bibitem{1} F.~Savart, 
%Memoires sur la constitution des veines liquides lancées par des orifices circulaires en mince paroi. 
Ann.\ de Chim.\ {\bf53}, 337 (1833).

\bibitem{2} Lord Rayleigh, 
%On the stability of jets. 
Proc.\ London Math.\ Soc.\ {\bf s1-10}, 4 (1878).

\bibitem{3} Lord Rayleigh, 
%On the instability of a cylinder of viscous liquid under capillary force.
Philos.\ Mag.\ 34, 145 (1892).

\bibitem{chandrasekhar} S.~Chandrasekhar, The capillary instability
of a liquid jet. {\em In Hydrodynamic and Hydromagnetic
Stability}, p.\ 537. Oxford Univ.\ Press (1961).

\bibitem{4} J.~Eggers, 
%Nonlinear dynamics and breakup of free-surface flows. 
Rev.\ of Mod.\ Phys.\ {\bf 69}, 865 (1997).

\bibitem{5} S.P.~Lin and R.D.~Reitz, 
%Drop and Spray Formation from a Liquid Jet. 
Annu.\ Rev.\ Fluid Mech.\ {\bf 30}, 85  (1998).

\bibitem{6} A.~Kalaaji, B.~Lopez, P.~Attane, and A.~Soucemarianadin,
%Breakup length of forced liquid jets. 
Phys.\ Fluids {\bf 15}, 2469 (2003).

\bibitem{7} O.A.~Basaran, 
%Small-scale free surface flows with breakup: Drop formation and emerging applications. 
AIChE J.\ {\bf 48}, 1842 (2002).

%\bibitem{8} S.~Biehl, R.~Danzebrink, P.~Oliveira, and M.A.~Aegerter,
%Refractive Microlens Fabrication by Ink-Jet Process. 
%J.\ of Sol-Gel Science and Technology {\bf 13}, 177  (1998).

%\bibitem{9} A.P.~Blanchard, R.J.~Kaiser, and L.E.~Hood,
%High-density oligonucleotide arrays.
%Biosensors and Bioelectronics {\bf 11}, 687 (1996).

\bibitem{10} B.A.M.~Hansson and H.M.~Hertz, 
%Liquid-jet laser-plasma extreme ultraviolet sources: from droplets to filaments. 
J.\ Phys.\ D {\bf 37}, 3233 (2004).

\bibitem{11} S.~Ter-Avetisyan, M.~Schn\"urer, P.V.~Nickles, M.~Kalashnikov, 
E.~Risse, T.~Sokollik, W.~Sandner, A.~Andreev, and V.~Tikhonchuk,
%Quasimonoenergetic Deuteron Bursts Produced by Ultraintense Laser Pulses. 
Phys.\ Rev.\ Lett.\ {\bf 96}, 145006 (2006).

\bibitem{12} \"O.~Nordhage, Z.-K.~Li, C.-J.~Friden, G.~Norman, U.~Wiedner, 
%On the behavior of micro-spheres in a hydrogen pellet target. 
Nucl.\ Instrum.\ Meth.\ A {\bf 546}, 391 (2005).

%\bibitem{13}  A.V.~Bucharov, M.~B\"uscher, A.F.~Ginevskiy, A.S.~Dmi\-triev, 
%V.P.~Chernyshov, V.D.~Chernetsky, 
%Pellet target for experiments on internal beam of accelerators. 
%Proc.\ $5^{\mathrm{th}}$ Ann.\ Conf.\ Cryogenics, Czech Republic, 
%      Praha, 1998, p.96.

%\bibitem{14}  A.V.~Bucharov, M.~B\"uscher, A.F.~Ginevskiy, A.S.~Dmi\-triev, 
%V.P.~Chernyshov, V.D.~Chernetsky,
%About heat-physical phenomena during generation and distribution of cryogenic corpuscular targets in cryo-technique. 
%Proc.\ $2^\mathrm{nd}$ Russian National Conf.\ on Heat Exchange. 
%Moscow 1998, v.~4, p.\ 280 (in Russian).

%\bibitem{15} A.V.~Bucharov, M.~B\"uscher, V.P.~Chernyshov, 
%Cryogenic corpuscular targets. The concept and the basic model. 
%MPEI Publishing, {\bf 17} (2002) (in Russian).

\bibitem{16} E.P.~Muntz and M.~Dixon,
%Applications to space operations of free-flying, controlled streams of liquids. 
J.\ Spacecraft Rockets {\bf 23}, 411 (1986).

\bibitem{non-linear} N.~Ashgriz, and F.~Mashayek, 
%Temporal analysis of capillary jet breakup, 
J.\ Fluid.\ Mech. {\bf 291}, 163 (1995);
R.~Suryo, P.~Doshi, and O.A.~Basaran,
%Nonlinear dynamics and breakup of compound jets,
Phys.\ Fluids {\bf 18}, 082107 (2006). 

\bibitem{20} V.V.~Blazhenkov, A.F.~Ginevskii, V.F.~Gunbin, A.S.\ Dmi\-triev, and S.I.~Tscheglov,
%Study of the transition regime of forced capillary breakup of a liquid jet. 
Fluid Dynamics {\bf 30}, 544  (1995).

\bibitem{17} V.V.~Blazhenkov, A.F.~Ginevskii, V.F.~Gunbin, A.S.\ Dmi\-triev, and A.I.~Motin,
%Monodisperse breakup of liquid jets. 
J.\ Eng.\ Phys.\ Thermophys.\ {\bf 55}, 994 (1988).

\bibitem{18} V.V.~Blazhenkov, A.F.~Ginevskii, V.F.~Gunbin, and A.S.\ Dmi\-triev,
%Forced capillary breakup of liquid jets. 
Fluid Dynamics {\bf 23}, 203  (1988); 
V.V.~Blazhenkov, A.F.~Ginevskii, V.F.~Gunbin, A.S.\ Dmi\-triev, and A.I.~Shcheglov,
%Nonlinear evolution of waves in forced decaying capillary jets. 
Fluid Dynamics {\bf 28}, 338  (1993).

%\bibitem{21} A.V.~Boukharov, A.A.~Semenov,
%Research on the behaviour of strongly evaporating capillary liquid jets in vacuum conditions. 
%Bulletin MPEI 4 2003, MPEI Publishing, 40 -- 42 (2003) (in Russian).

%\bibitem{22} C.~Weber,
%Zum Zerfall eines Flüssigkeitsstrahles. 
%Z.\ Angew.\ Math.\ Mech.\ {\bf 11}, 136 (1931).

%\bibitem{23} A.S.~Sterling, and C.A.~Sleicher,
%The instability of capillary jets. 
%J.\ Fluid Mech.\ {\bf 68}, 477 -- 495 (1975).

\bibitem{24} W.T.~Pimbley, and H.C.~Lee,
%Satellite Droplet Formation in a Liquid Jet. 
IBM J.\ Res.\ Dev.\ {\bf 21}, 21 (1977).

\bibitem{27} \"O.~Nordhage,
%On a hydrogen pellet target for antiproton physics with PANDA (Ph.D. Thesis). 
Ph.D.\ Thesis, Acta Universitatis Upsaliensis (2006);\hfill\\
http://publications.uu.se/abstract.xsql?dbid=7137

\bibitem{patent} 
Russian Federation Patent No.\ 2298890;
German Patent Application No.\ 10\,2007\,017\,212.7-13

\bibitem{25} D.U.~Hamburg and N.F.~Dubrovkin, Hydrogen: properties,
  generation, storage, transport, application, Chemistry Publishing,
  Moscow (1989), ISBN 5-7245-0034-5, pp.\ 109, 142, 174 (in Russian).

\bibitem{26} M.P.~Malkov (ed.), Handbook of physical and technical
  fundamentals of cryogenics, Energoatomizdat Publishing, Moscow
  (1963) pp.\ 90, 123, 176 (in Russian).

\bibitem{evaporation} M.~Saroka, Y.~Guo, and N.~Ashgriz,
%Nonlinear instability of an evaporating capillary jet, 
AIAA J.\ {\bf 39}, 1728 (2001); 
S.~Savtchenko, and N.~Ashgriz,
%Temporal instability of a capillary jet with a source of mass,
Phys.\ Fluids {\bf 17}, 112102 (2005).

\bibitem{deflection} J.M.~Chwalek {\em et al.}, 
%D.P.~Trauernicht, C.N.~Delametter,
%  R.~Sharma, D.L.~Jeanmaire, C.N.~Anagnostopoulos, G.A.~Hawkins,
%  B.~Ambravaneswaran, J.C.~Panditaratne, O.A.~Basaran, 
%A new method for deflecting liquid microjets, 
Phys.\ Fluids {\bf 14}, L37 (2002).


\end{thebibliography}
\end{document}